\documentstyle[12pt]{article}
\newcounter{mycount}

\newcommand{\bee}{\begin{eqnarray}}
\newcommand{\eee}{\end{eqnarray}}
\newcommand{\be}{\begin{eqnarray}}
\newcommand{\ee}{\end{eqnarray}}
\newcommand{\ddxi}{\frac {\mbox{d}} {\mbox{d}\xi} }
\newcommand{\CH}{\frac {\mbox{ch}(3\nu\xi)} {3\nu}}
\newcommand{\SH}{\mbox{sh}(3\nu\xi)}

\newcommand\nn{\nonumber \\}
\newcommand\ch{\mbox{ch}}
\newcommand\sh{\mbox{sh}}
\newcommand{\p}{\partial}
\renewcommand{\forall}{\mbox{ for any }}
\font\frtnfr=eufm10   scaled\magstep1
\font\twlfr=eufm10
\font\tenfr=eufm10

\newfam\frfam
\textfont\frfam=\frtnfr
\scriptfont\frfam=\twlfr
\scriptscriptfont\frfam=\tenfr
\def\fr{\fam\frfam}

\font\frtnopen=msbm10  scaled\magstep2
\font\twlopen=msbm10
\font\tenopen=msbm10

\newfam\openfam
\textfont\openfam=\frtnopen
\scriptfont\openfam=\twlopen
\scriptscriptfont\openfam=\tenopen
\def\open{\fam\openfam}

\font\frtnsf = cmss12 scaled\magstep1
\font\twlsf = cmss10
\font\tensf = cmss9

\newfam\Scfam
\textfont\Scfam = \frtnsf
\scriptfont\Scfam = \twlsf
\scriptscriptfont\Scfam = \tensf

\begin{document}
\renewcommand{\theequation}{\arabic{equation}}
\bibliographystyle{nphys}
\setcounter{equation}{0}
\sloppy
\title
 {
      \hfill{\normalsize\sf FIAN/TD/97-17}    \\
            \vspace{1cm}
3-particle Calogero Model:
Supertraces and Ideals on the Algebra of  Observables
 }
\author
 {
              S.E.Konstein
          \thanks
             {E-mail: konstein@td.lpi.ac.ru}
          \thanks
             {
               This work was supported
               by the Russian Basic Research Foundation, grant 96-01-01144
               and grant 96-15-96463.
             }
  \\
               {\small \phantom{uuu}}
  \\
               {\sf \small I.E.Tamm Department of Theoretical Physics,}
  \\
               {\sf \small P. N. Lebedev Physical Institute,}
  \\
               {\sf \small 117924, Leninsky Prospect 53, Moscow, Russia.}
 }
\date{ }
\maketitle

\renewcommand{\theequation}{\arabic{equation}}
\setcounter{equation}{0}
\begin{abstract}
{
The associative superalgebra of observables of 3-particle Calogero model
giving all wavefunctions of the model via standard Fock procedure
has 2 independent supertraces. It is shown here that when
the coupling constant $\nu$ is equal to
$n+1/3$, $n-1/3$ or $n+1/2$
for any integer $n$
the existence of 2 independent supertraces
leads to existence of nontrivial two-sided ideal in the
superalgebra of observables.
}
\end{abstract}


\section{Introduction}

Three-particle Calogero model is a convenient object for investigation
of some properties of $N$-particle Calogero models \cite{4}, as it has
many features which distinguish multiparticle systems from one-particle ones,
but still remains an easy problem admitting separation of variables
\cite{OP}.

It is well known that after some similarity transformation the Hamiltonian
of Calogero model attains the form
\be\label{Cal}
H_{Cal} =-\frac 1 2
\sum_{i=1}^N \left[ \frac {\p^2} {\p x_i^2} -x_i^2
+ \nu \sum_{j\neq i} \frac 2 {x_i - x_j}
\frac {\p} {\p x_i} \right]
\ee
which coincides with the operator
\be\label{Ca}
H=\frac 1 2 \sum_{i=1}^N \left\{a^0_i,\, a^1_i\right\}
\ee
on the space of symmetric functions.
In this expression the annihilation and creation operators
 $a^\alpha_i$ ($\alpha=0,\,1$) are expressed through the
differential-difference Dunkl operators  $D_i(x)$ \cite{6}
\bee\label{dunkl}
a^\alpha_i= \frac 1 {\sqrt{2}} (x_i +(-1)^\alpha D_i(x)),
\mbox { where }
D_i= \frac \p {\p x_i}+
\nu \sum_{l\neq i}^N  \frac {1} {x_i-x_l}(1-K_{il}),
\eee
and $K_{ij}=K_{ji}$ are the operators of elementary transpositions
\be
K_{ij} x_i = x_j K_{ij},\quad K_{ij}x_k=x_k K_{ij} \mbox{ if }k\neq i
\mbox{ and } k\neq j.
\ee
The operators  $a^\alpha_i$ satisfy the following commutation relations
\bee\label{AA}
\left [ a^\alpha_i\,,\,a^\beta_j \right ]= \epsilon^{\alpha\,\beta}
\left ( \delta_{ij} +
\nu\delta_{ij}\sum_{l=1,\,l\neq i}^N K_{il}-\nu\delta_{i\neq j}K_{ij}
\right),\nonumber
\eee
where
$\epsilon^{\alpha\,\beta}$ = $- \epsilon^{\beta\,\alpha},$
$\epsilon^{0\,1}=1$, which lead (\cite{2}) to the relations
\be
\left [H,\, a^\alpha_i \right ] = - (-1)^\alpha a^\alpha_i ,\nonumber
\ee
and, therefore, $a_i^1$ when acting on the Fock vacuum $|0\rangle$
such that $a_j^0|0\rangle = 0$ for any $j$,
give all the wavefunctions of the model (\ref{Ca})
and, consequently, of the model (\ref{Cal}). In \cite{3}
Perelomov operators \cite{Per},
describing the wavefunctions of the model (\ref{Cal}) are expressed
via these operators $a^\alpha_i$.

Let $SH_N(\nu)$ be the associative superalgebra of all polynomials in the
operators  $K_{ij}$   and
$a^\alpha_i$  (\ref{dunkl}), with
 ${\open Z}_2$-grading $\pi$ defined by the formula:
$\pi(a^\alpha_i)=1$, $\pi(1)=\pi(K_{ij})=0$.
The superalgebra $SH_N(\nu)$ is called
{\it the algebra of observables}.
It is clear that $SH_N(\nu) = SH_1(0)\otimes SH_N^\prime(\nu)$,
where $SH_1(0)$ is generated by the operators
$a^\alpha = \sum_i a^\alpha_i/\sqrt N$
satisfying the commutation relations
$[a^\alpha, a^\beta] =\epsilon^{\alpha\beta}$,
while $SH_N^\prime(\nu)$ is generated by the transpositions $K_{ij}$
and by all linear combinations of the form
$\sum_i \lambda_i a^\alpha_i$ with coefficients $\lambda_i$
satisfying the condition
$\sum_i\lambda_i=0$.

The group algebra
${\open C}[S_N]$ of the permutation group $S_N$,
generated by elementary
transpositions $K_{ij}$ is a  subalgebra of $SH_N^\prime(\nu)$.

The group $S_N$
acts trivially on the space of wavefunctions of Calogero model
(\ref{Cal}),
but for every nontrivial representation of $S_N$
there exists a matrix version of the Calogero model
where this representation is realized
\cite{VDK}.

The {\it supertrace} on an arbitrary associative superalgebra ${\cal A}$
is  complex-valued linear function
$str(\cdot)$
on ${\cal A}$
which satisfies the condition
\be\label{str}
str(fg)=(-1)^{\pi(f)\pi(g)} str(gf)
\ee
for every $f,g \in {\cal A}$ with definite parity.

It is proven in \cite{kv} that $SH_N(\nu)$  has nontrivial supertraces and
the dimension of the space of the supertraces on $SH_N(\nu)$
is equal to the number of partitions of $N$
into the sum of positive odd integers.
In particular,
$SH_2(\nu)$
has only one supertrace and
$SH_3(\nu)$
has two independent supertraces. Since
$SH_N(\nu) = SH_1(0)\otimes SH_N^\prime(\nu)$
and $SH_1(0)$ has one supertrace, it follows that
the numbers of supertraces on $SH_N(\nu)$ and on $SH_N^\prime(\nu)$
are the same.

Knowing the supertraces on
$SH_N^\prime(\nu)$
is useful in different respects.
One of the most important applications is that
these define the multilinear invariant forms
$$
str(f_1 f_2\,...\,f_n),
$$
which makes it possible, for example, to construct
the Lagrangians of dynamical theories
based on these algebras.
In particular, every supertrace defines
the bilinear form
\be\label{form}
B_{str}(f,\,g)=str(fg).
\ee
Since the complete set of null-vectors
of every invariant bilinear form constitutes
the two-sided ideal of the algebra,
the supertrace is a good tool for finding
such ideals.
In the well known case of $SH_2^\prime(\nu)$
corresponding to the usual two-particle Calogero model
the only supertrace defines the bilinear form
(\ref{form}) which degenerates
when $\nu$ is half-integer
\cite{14}, hence, the superalgebra
$SH_2^\prime(\nu)$
has an ideal for these values of
$\nu$.
When
$N\ge 3$
the situation is more complicated
as $SH_N^\prime (\nu)$ has more than one supertrace.
In the case of finitedimensional
superalgebra this would be sufficient for existence of ideals
but for infinitedimensional superalgebras under consideration
the existence of ideals is not completely
investigated yet.

Here I show
that in the superalgebra
$SH_3^\prime (\nu)$ with
$\nu=n\pm 1/3$
or
$\nu=n+1/2$
there exists one supertrace with degenerated
corresponding bilinear form (\ref{form}) and
that for the other values of $\nu$
all the bilinear forms corresponding to supertraces are non-degenerate.
It should be noticed that the existence of supertraces
depends on the choice of
${\open Z}_2$-grading
while the existence of an ideal does not depend on the grading.
Still open are the problems, what the quotient algebras are
(in the case of
$SH_2(\nu)$
for
$\nu=n+1/2$
these are
$Mat_n$) and if there exist ideals of another type.

The paper is organized as follows. In Section 2
the generating elements of associative superalgebra
$SH_3^\prime(\nu)$
are presented together with the relations between them.
The equation for generating functions of arbitrary supertrace
on
$SH_3^\prime(\nu)$
are derived and solved in  Section 3,
and in the last Section all values of $\nu$,
such that
$SH_3^\prime(\nu)$
has the ideal generated by the zeroes of the bilinear form
corresponding to the supertrace are found.

\section{The algebra
$SH_3^\prime(\nu)$}
In this Section we present the generating elements of associative superalgebra
$SH_3^\prime(\nu)$ and relations between them.

Let
$\lambda = exp(2\pi i/3)$.
       As a basis in
${\open C}\left [ S_3 \right ]$ let us choose
the vectors
\bee
L_k &=&  \frac 1 3 (\lambda^k K_{12}+K_{23}+\lambda^{-k} K_{31}), \nn
Q_k &=& \frac 1 3 (1 + \lambda^k K_{12}K_{13} + \lambda^{-k} K_{12}K_{23}),\nn
&{}&L_\pm\stackrel {def}{=}  L_{\pm 1},\quad
Q_\pm \stackrel {def}{=} Q_{\pm 1}. \nonumber
\eee
          Instead of the generating elements
$a^\alpha_i$ let us introduce
the vectors
\bee
x^\alpha &=& a^\alpha_1+ \lambda a^\alpha_2+ \lambda^2 a^\alpha_3,
\ \ \ x \stackrel {def}{=} x^0, \ \ \ x^+ \stackrel {def}{=} x^1,   \nn
y^\alpha &=& a^\alpha_1+ \lambda^2 a^\alpha_2+ \lambda a^\alpha_3,
\ \ \ y \stackrel {def}{=} y^0, \ \ \ y^+ \stackrel {def}{=} y^1,  \nonumber
\eee
which are derived from
$a^\alpha_i$
by subtracting the center of mass
$a^\alpha =
\displaystyle{\frac 1 3 \sum_{i=1,2,3} a^\alpha_i}$.

The Lie algebra
${\fr sl}_2$
of inner automorphisms of
$SH_3^\prime(\nu)$
is generated by the generators
$T^{\alpha \beta}$
\bee\label{sl2}
T^{\alpha \beta} = \frac 1 3 (x^\alpha y^\beta + x^\beta y^\alpha).
\ee
These generators satisfy the usual commutation relations
\be \label {csl2}
        [T^{\alpha\beta}, T^{\gamma\delta}] =
   \epsilon^{\alpha\gamma} T^{\beta\delta}  +
   \epsilon^{\alpha\delta} T^{\beta\gamma}  +
   \epsilon^{\beta\gamma}  T^{\alpha\delta} +
   \epsilon^{\beta\delta}  T^{\alpha\gamma} \,,
\ee
and act on generating elements
$x^\alpha$
and
$y^\alpha$
as follows:
\be \label {sl2vec}
\left [ T^{\alpha\beta},\,x^\gamma \right ]=\epsilon^{\alpha\gamma} x^\beta
                                  +\epsilon^{\beta\gamma} x^\alpha \,,\qquad
\left [ T^{\alpha\beta},\,y^\gamma \right ]=\epsilon^{\alpha\gamma} y^\beta
                                  +\epsilon^{\beta\gamma} y^\alpha
\ee
leaving the group algebra ${\open C}[S_3]$ invariant:
$\left[ T^{\alpha\beta},\, K_{ij}\right]=0$.

Clearly, $SH_3^\prime(\nu)$
decomposes into the infinite direct sum of finitedimensional
irreducible representations of this ${\fr sl}_2$.
Every subspace of the space $SH_3^\prime(\nu)$ where
${\fr sl}_2$ (\ref{sl2}) acts non-trivially and irreducibly
consists of linear combinations of the vectors of the form
$[f,\,T^{\alpha\beta}]$. So every supertrace vanishes on this
subspace and has non-trivial values only on the associative
subalgebra $H_3^0(\nu) \subset SH_3^\prime(\nu)$ of all ${\fr sl}_2$-singlets.

Let
$m=\frac 1 4 \left \{ x^\alpha,\, y_\alpha\right\}$.
Clearly, $m$ is a singlet under
the              action of ${\fr sl}_2$
(\ref{sl2});
$m$ can also be expressed in the form
\bee
\label{mx}
m &=& \frac 1 2 \left  ( x^\alpha y_\alpha +3 +9\nu L_0 \right)
\mbox{   or, equivalently,}\\
\label{my}
m &=& \frac 1 2 \left  ( y_\alpha x^\alpha -3 -9\nu L_0 \right).
\eee
In these formulas the greek indices are lowered and rised
with the help of the antisymmetric tensor $\varepsilon^{\alpha\beta}$:
$a^{\alpha}= \sum_\beta \varepsilon^{\alpha\beta}a_\beta $.
Obviously, all the other singlets in
$H_3^0(\nu)$ are polynomials in $m$ with coefficients in
${\open C}[S_3]$.

The generating elements
$x^\alpha$,
$y^\alpha$,
$Q_i$,
$L_i$ and the element $m$
satisfy the following relations:
\bee
\label{LL}
L_i L_j = \delta_{i+j} Q_j , \qquad
\label{LQ}
L_i Q_j = \delta_{i-j} L_j , \\
\label{QL}
Q_i L_j = \delta_{i+j} L_j ,  \qquad
\label{QQ}
Q_i Q_j = \delta_{i-j} Q_j ,
\eee
where $\delta_i=\delta_{i\, 0}$,
\bee
\label{Lx}
L_i x^\alpha = y^\alpha L_{i+1},  & \qquad &
\label{Ly}
L_i y^\alpha = x^\alpha L_{i-1},   \\
\label{Qx}
Q_i x^\alpha = x^\alpha Q_{i+1},  & \qquad &
\label{Qy}
Q_i y^\alpha = y^\alpha Q_{i-1},   \\
\label{Lm}
L_i m = -m L_i                 ,  & \qquad &
\label{Qm}
Q_i m =  m Q_i                 ,
\eee
\bee
\left [ x, \, x^+ \right ]  =-9\nu L_{+1} ,\qquad
\left [ y, \, y^+  \right ] =-9\nu L_{-1},\qquad
\left [ y, \, x^+  \right ] =\left [ x, \, y^+ \right ] = 3+9\nu L_0,
\eee
and
\bee
\left [ m, \, x^\alpha  \right ] &=&\frac 3 2 \left ( x^\alpha
           + 3 \nu x^\alpha L_0 +   3 \nu L_0 x^\alpha \right ),         \nn
\left [ m, \, y^\alpha \right ] &=& - \frac 3 2 \left ( y^\alpha
           + 3 \nu y^\alpha L_0 +   3 \nu L_0 y^\alpha \right ),
\eee
from which the relations
\bee
\label{mx-}
mQ_+ x^\alpha Q_- &=& Q_+ x^\alpha Q_- \left (m+\frac 3 2 \right ) ,   \\
\label{my+}
mQ_- y^\alpha Q_+ &=& Q_- y^\alpha Q_+ \left (m-\frac 3 2 \right ) ,   \\
\label{mx0}
mQ_- x^\alpha Q_0 &=& Q_- x^\alpha Q_0 \left (m+\frac 3 2
                                    + \frac 9 2 \nu L_0 \right )   ,   \\
\label{my0}
mQ_+ y^\alpha Q_0 &=& Q_+ y^\alpha Q_0 \left (m-\frac 3 2
                                    - \frac 9 2 \nu L_0 \right )   ,   \\
Q_0 x^\alpha Q_+ m &=& \left (m-\frac 3 2
                   - \frac 9 2 \nu L_0 \right ) Q_0 x^\alpha Q_+   ,   \\
Q_0 y^\alpha Q_- m &=&  \left (m+\frac 3 2
                   + \frac 9 2 \nu L_0 \right ) Q_0 y^\alpha Q_-
\eee
follow.

It was shown in \cite{kv} that every supertrace on
$SH_3^\prime(\nu)$
is completely determined
by its values on
${\open C}\left [ S_3 \right ] \subset SH_3^\prime(\nu)$,
 i.e.,   by the values
$str(1)$, $str(K_{12})$ = $str(K_{23})$ = $str(K_{31})$
and
$str(K_{12}K_{23}) = str(K_{12}K_{13})$,
which are consistent with (\ref{str}) and
admit the extension of the supertrace from
${\open C}\left [ S_3 \right ]$
to
$SH_3^\prime(\nu)$ in a unique way, if and only if
\be\label{strL0}
str(K_{ij})= \nu (-2str(1) - str(K_{12}K_{23})).
\ee

It was noticed above that
only trivial
representations of ${\fr sl}_2$ can contribute to any supertrace on
$SH_3^\prime(\nu)$.
Therefore, it suffices to find the restrictions of the supertraces
on $H_3^0(\nu)$.

\section{Generating functions}
The equations for generating functions
of arbitrary supertrace are derived and solved in this section.

Let us introduce {\it the generating functions}:
\bee
\label{F}
F(\xi) &=& str \left( \sh\left( \frac 2 3 \xi m\right)\left(Q_+ - Q_-\right)
                                                       \right )\,,   \\
\label{Phi}
\Phi(\xi)&=& str \left( \ch\left( \frac 2 3 \xi m\right)\left(Q_+ + Q_-\right)
                                                        \right )\,,  \\
\label{Psi}
\Psi(\xi)&=& str \left( \ch\left( \xi
                 \sqrt {\frac 4 9 m^2 + 9 \nu^2} \right) Q_0 \right ),
\eee
which completely describe any supertrace $str(\cdot)$ on
$SH_3^\prime(\nu)$.
Indeed, it follows from
(\ref{Lm}) and (\ref{str})
that the identities
$str(m^k L_i)=str(m^{k-1}L_i m)=-str(m^k L_i)$
take place
for
$k>0$
and, therefore, $str(f(m)L_i)=f(0)str(L_i)$.
Further on, from
$str(m^k Q_0)=str(m^k L_0^2)=str(L_0 m^k L_0)
=(-1)^k str(m^k L_0^2)$ it follows that
$
str(\sh (\xi m) Q_0)=0.
$
Analogously,
$str(\sh (2/3 \xi m) (Q_+ + Q_-))=0$
and
$str(\ch (2/3 \xi m) (Q_+ - Q_-))=0$.

\vskip 3mm
{\bf The equations for generating functions.}

One can obtain two equations for generating functions of supertrace
by differentiating the definition
(\ref{F})
and two more equations by differentiating (\ref{Phi}).
To derive the first equation one has to use
the expression (\ref{mx}) for $m$
in the term
$mQ_+$
and
(\ref{my}) in the term
$mQ_-$.
To derive the second equation one has to use
the expression (\ref{my}) for $m$
in the term
$mQ_+$
and
(\ref{mx}) in the term
$mQ_-$.
The first equation has the following form
\bee
\label{F10}
F^\prime (\xi)=\frac 2 3
             str \left( \ch\left( \frac 2 3 \xi m\right)\left(
\frac 1 2 \left  ( x^\alpha y_\alpha +3 +9\nu L_0 \right) Q_+
-  \frac 1 2 \left  ( y_\alpha x^\alpha -3 -9\nu L_0 \right) Q_-\right)
                   \right )\,, \nonumber
\eee
which with the help of
(\ref{Phi}), (\ref{LQ}), (\ref{Qx}), (\ref{Qy}), (\ref{mx-}), (\ref{my+})
and the defining properties of supertraces (\ref{str})
can be reduced to
\bee
\label{F11}
F^\prime (\xi)=\Phi(\xi) - \frac 1 3
             str \left( \ch\left( \frac 2 3 \xi m +\xi \right)
y_\alpha x^\alpha Q_-
- \ch\left( \frac 2 3 \xi m -\xi \right) x^\alpha y_\alpha Q_+\right).
\eee
By ubstituting in (\ref{F11}) the expressions for
$x^\alpha y_\alpha$
and
$y_\alpha x^\alpha$  in terms of
$m$  obtained
from (\ref{mx}) and (\ref{my}), respectively, and
decomposing the expressions
$\ch\left( \frac 2 3 \xi m \pm \xi \right)$,
one can easily obtain the equation desired:
\bee
\label{F1}
\Phi=F^\prime + (\sh\xi\,\Phi)^\prime -(\ch\xi \, F)^\prime.
\eee
To derive the second equation we start with
\bee
\label{F20}
F^\prime (\xi)=\frac 2 3
             str \left(
                        \ch\left( \frac 2 3 \xi m\right)       \left(
    \frac 1 2 \left  ( y_\alpha x^\alpha -3 -9\nu L_0  \right) Q_+
 -  \frac 1 2 \left  ( x^\alpha y_\alpha +3 +9\nu L_0  \right) Q_-
                 \right)
                   \right ).          \nonumber
\eee
With the help of (\ref{my0}), (\ref{mx0}), (\ref{LQ}),
and (\ref{str}) one obtains
\bee
\label{F21}
F^\prime (\xi)&=&-\Phi(\xi)
- \frac 1 3  str \left(
    \ch\left( \frac 2 3 \xi m -\xi -3 \nu \xi L_0 \right)
                                                   x^\alpha y_\alpha Q_0
  - \ch\left( \frac 2 3 \xi m +\xi +3 \nu \xi L_0 \right)
                                                   y_\alpha x^\alpha Q_0
                 \right)\nn
{}&=&
-\Phi -\frac 1 3
             str \Bigg(   
 \ch\xi\, \ch\left( \frac 2 3 \xi m -3 \nu \xi L_0 \right)
                                                    (2m-3-9\nu L_0) Q_0
\nn
&{}&
-
\sh\xi\, \sh\left( \frac 2 3 \xi m -3 \nu \xi L_0 \right)
                                                    (2m-3-9\nu L_0) Q_0
\nn
&{}&
-
\ch\xi\, \ch\left( \frac 2 3 \xi m +3 \nu \xi L_0 \right)
                                                    (2m+3+9\nu L_0) Q_0\nn
&{}&-
\sh\xi\, \sh\left( \frac 2 3 \xi m +3 \nu \xi L_0 \right)
                                                    (2m+3+9\nu L_0) Q_0
                 \Bigg).     
\eee
{}From
$mL_0=-L_0 m$
it follows that
\bee
\label{ch}
\ch(\alpha m+\beta L_0)Q_0&=&\left(\ch\sqrt{\alpha^2 m^2+\beta^2}
                                                         \right)Q_0\,,\nn
\label{sh}
\sh(\alpha m+\beta L_0)Q_0&=&\frac 1 {\sqrt{\alpha^2 m^2+\beta^2}}
       \left(\sh\sqrt{\alpha^2 m^2+\beta^2}\right)(\alpha m+\beta L_0)Q_0,
\nonumber
\eee
which allows us to rewrite equation (\ref{F21})
in the following form
\bee
\label{F2}
F^\prime = -\Phi
  + 2 \left( \sh \xi str \left(\ch\sqrt{(\frac 2 3 \xi m)^2
            +(3\nu\xi)^2}\right) Q_0 \right)^\prime
  +2 (\ch \xi\, \sh(3\nu\xi) \,str L_0)^\prime.   \nonumber
\eee

The second pair of equations is derived in a similar way.

Finally, one obtains the following
system of equations
\bee
\label{1_}
F^\prime = -\Phi
  + 2 \left( \sh \xi \Psi \right)^\prime
  +2 (\ch \xi\, \sh(3\nu\xi) \,str L_0)^\prime  \,, \\
\label{2_}
\Phi^\prime = -F
  - 2 \left( \ch \xi \Psi \right)^\prime
  - 2 (\sh \xi\, \sh(3\nu\xi) \,str L_0)^\prime  \,, \\
\label{3_}
\Phi=F^\prime + (\sh\xi\,\Phi)^\prime -(\ch\xi \, F)^\prime \,, \\
\label{4_}
F=\Phi^\prime + (\ch\xi\,\Phi)^\prime -(\sh\xi \, F)^\prime \,,
\eee
(where the equations (\ref{3_}) and (\ref{4_}) are equivalent)
with the following initial conditions
\bee
\label{IF}
            F(0)&=&0             \,,    \nn
\label{IPhi}
            \Phi(0)&=&str(Q_++Q_-)\,,   \nn
\label{IPsi}
            \Psi(0)&=&str(Q_0)    \,.   \nonumber
\eee

It is easy to find three integrals of motion
for this system.
To find the first one, it suffices to
subtract (\ref{3_}) from (\ref{1_}),
to find the second one, to subtract (\ref{4_}) from (\ref{2_}).
To find the third one, one considers
the linear combination of equations
$\sh\xi \,$(\ref{1_})+$\ch\xi\,$(\ref{2_}).

The result of these actions has
the following form:
\bee
\left(
F\,(2-\ch\xi) + \Phi\, \sh\xi + \Psi\, (-2\sh\xi)
\right)^\prime
&=& 2 (\ch\xi\,\sh(3\nu\xi)str(L_0) )^\prime   \,,\nn
\left(
F\,(-\sh\xi) + \Phi\, (2+\ch\xi) + \Psi\, (2\ch\xi)
\right)^\prime
&=& -2 (\sh\xi\,\sh(3\nu\xi)str(L_0) )^\prime   \,,\nn
\left(
F\,\sh\xi + \Phi\, \ch\xi + 2 \Psi
\right)^\prime
&=& -2 \sh(3\nu\xi)str(L_0)                     \,. \nonumber
\eee

\vskip 3mm
{\bf The solution}.

The solution of the obtained system of equations in the terms of
generating functions  more convenient for further consideration
is of the form
\bee
\label{+}
    {\cal Q}_+ \stackrel {def}{=} str (e^{2/3\xi m} Q_{+ 1}) =
    \frac {P_+}{\Delta}, \mbox{ where \hskip 8.2 cm}\nn
    P_+ =
     {2\nu S_1} \left(
       \CH \left( -e^{ 2\xi} + 2 e^{- \xi}\right)
       +
       \SH \left( e^{ 2\xi} +  e^{- \xi}\right)
     \right) +
 \frac {S_2} {2} \left( e^{- 2\xi} -2 e^{ \xi} + 3 \right),\\
\label{-}
    {\cal Q}_- \stackrel {def}{=} str (e^{2/3\xi m} Q_{- 1}) =
    \frac {P_-}{\Delta}, \mbox{ where \hskip 8.2 cm}\nn
    P_- =
     {2\nu S_1}
     \left(
     \CH \left( -e^{-2\xi} + 2 e^{\xi}\right)
     -
     \SH \left( e^{- 2\xi} +  e^{ \xi}\right)
     \right)
     + \frac {S_2} {2}
     \left( e^{ 2\xi} -2 e^{-\xi} + 3 \right), \\
\label{0}
     {\cal Q}_{0}^\pm \stackrel {def}{=}
      str (\ch(\xi \sqrt{4/9 m^2 +9\nu^2} (Q_{0}\pm L_0))=
    \frac {P_0}{\Delta} \pm \nu S_1 \ch (3\nu \xi), \mbox{ where
                                      \hskip 2.35 cm}\nn
      P_0 = {2\nu S_1}
      \left(-3\CH + \frac {\SH} 2
             \left(- e^{3\xi} +  e^{-3\xi}\right)\right)
     + \frac {S_2} {2}
     \left( e^{2\xi} + e^{-2\xi} - 2 e^{\xi} -2 e^{-\xi}  \right),
\eee
$${\cal R}_{0} \stackrel {def}{=} str (sh(\xi m) Q_{0})=0 ,$$
$${\cal L}_{\pm} \stackrel {def}{=} str (e^{2/3\xi m} L_{\pm 1})=0 ,$$
$${\cal L}_{0} \stackrel {def}{=} str (e^{2/3\xi m} L_{0})=\nu S_1 .$$
Here
$\Delta=exp(-3\xi)(exp(3\xi)+1)^2$
and
$S_1$, $S_2$  are arbitrary parameters determining the supertrace
in the two-dimensional space
of supertraces:
\bee
S_1=-2str(1)-str(K_{12}K_{23}),\qquad
S_2=\frac 8 3 str(1) -\frac 2 3 str(K_{12}K_{23}) .  \nonumber
\eee

\section{Ideals}
\noindent
It can be easily shown that if some non-trivial
ideal ${\cal I}\subset SH^\prime_3(\nu)$
has nonzero intersection with $H_3^0(\nu)$ (which is an ideal
in $H_3^0(\nu)$),
then ${\cal I}$ has nonzero intersection at least with one of the subspaces
in $H_3^0(\nu)$
consisting either of the elements of the form
$\{ Q_\pm  f(m) \}$ (cases $P_\pm$)
or of the elements of the form
$\{(Q_0 \pm L_0 ) f(m) \}$ (case $P_0$).

{}First, let us look for elements of the ideal of the form
$\{ Q_+  f(m) \}$,  i.e.,
look for polynomial functions
$f$ such that
\bee
    f\left( \ddxi \right){\cal Q}_+ =0.   \nonumber
\eee
If such nontrivial function $f$ does exist,
then (\ref{+}) is a linear combination of exponents
\bee\label{++}
{\cal Q}_+=\sum_i p_i e^{b_i\xi}   \,,\nonumber
\eee
where each
$p_i$
is a polynomial in $\xi$.

It follows from (\ref{+}) that
$p_i$'s are constants and that the above presentation
exists if and only if
all zeroes of the function
$\Delta(\xi)$
are  zeroes of
$P_+(\xi)$,  i.e.,
\bee\label{1}
P_+((2k+1)\pi i/3)=0 \qquad \forall k\in {\open Z},      \\
\label{2}
P_+{}' ((2k+1)\pi i/3)=0 \qquad \forall k\in {\open Z}.
\eee
The second condition reflects the fact that all zeroes
of the function $\Delta$ are of multiplicity 2.

Conditions (\ref{2}) are  identically satisfied
for all values of
$S_1$, $S_2$ and $\nu$.
The conditions (\ref{1}) lead to the system of equations
\bee\label{sys}
\frac {S_1} 3 \cos((2k+1)\pi\nu) + \frac {S_2} 2 \cos((2k+1)\pi/3) =0
\qquad \forall k\in {\open Z},
\eee
which has nontrivial solutions for $S_1$ and $S_2$
if and only if
\bee\label{KL}
\cos((2k+1)\pi\nu) \cos((2l+1)\pi/3) = \cos((2l+1)\pi\nu) \cos((2k+1)\pi/3)
\nonumber
\eee
for any integer $k$ and $l$.
The case $l=3$, $k=2$ gives
$\cos(7\pi\nu) - \cos(5\pi\nu) = 0$
and, therefore,
$\nu=s/6$, where
$s$ is an integer.
To determine $s$, consider two cases:
$k,l$ and $k+3$, $l$.
The immediate consequence of these cases is
\bee
(\cos(s\pi)-1)\cos((2k+1)s\pi/6)=0 \qquad \forall k \in {\open Z}.  \nonumber
\eee
Hence, either
$s=2n$, or
for any $k \in {\open Z}$ there exists $n \in {\open Z}$,
such that $(2k+1)s=6n+3$  i.e.,  $s=6l+3$
for an integer $l$.

Finally,
the subspace consisting of the elements of the form
$\{ Q_+  f(m) \}$
(case $P_+$)
is nontrivial in ${\cal I}$ if and only if
\bee\label{otwet}
\nu &=& n \pm \frac 1 3 \,, \qquad \frac {S_1} 3 = -(-1)^n \frac {S_2} 2
           \mbox{\hskip 3mm  or}\nn
\nu &=& n + \frac 1 2\,, \qquad S_2=0
\eee
with         $n$ integer.

The cases $P_-$ and $P_0$ give the same values of
$\nu$, $S_1$ and $S_2$.

\vskip 3mm
{\bf The other values of
$\nu$}.

It is proved above that the subalgebra of singlets
$H_3^0(\nu)\subset SH_3^\prime(\nu)$
contains an ideal generated by a supertrace
if and only if
$\nu=n\pm 1/3$
or
$\nu=n+1/2$. If
$H_3^0(\nu)$
contains such an ideal the latter can be extended
to a nontrivial ideal in
$SH_3^\prime(\nu)$ in an obvious way.
It will be proved in this section that if
all bilinear forms generated by supertraces
$str(f\cdot g)$
are non-degenerate on the subalgebra
$H_3^0(\nu)$, then they are non-degenerate on the superalgebra
$SH_3^\prime(\nu)$ too,  i.e.,  that for all the other values
of
$\nu$
there are no ideals generated by supertraces in the superalgebra
$SH_3^\prime(\nu)$.

To this end it suffices to prove
that none of the following systems of equations
\be\label{ur}
\sum_{j=0}^n M^n_{ij} p_j =0, \ \ i=1,\,...\,,n
\ee
for
$p_j \in H_3^0(\nu)$
has non-trivial solutions.
The matrix elements
$M^n$
in system (\ref{ur}) are
\be
M^n_{ij} = \left( x^i y^{n-i} (x^+)^{n-j} (y^+)^j \right)_0,
\ee
where the subscript $0$ singles out
the ${\fr sl}_2$-singlet part
$(f)_0$
from the polynomial $f$.

Clearly, the elements of the matrix
$M^n$
are the polynomial in
$m$
of degree not greater than
$n$
with coefficients in
${\open C}[S_3]$:
$M^n_{ij}=\sum_{k=0}^n c_{ij}^k m^k$,
$c_{ij}^k \in {\open C}[S_3]$,
$c_{ij}^n \in {\open C}$.
It is also obvious that non-degeneracy
of the matrix
${\tilde M}^n$
with elements
$({\tilde M}^n)_{ij}=c_{ij}^n$
leads to the absence
of nontrivial solutions
of (\ref{ur}).

Let us find the coefficients
$c_{ij}^n$. Since,
up to polynomials in
$x$
and
$y$
of lesser degrees,
$m^n$
is a linear combination of the
monomials of the form
$x^i y^{n-i} (x^+)^{n-i} (y^+)^i$, it follows that
$c_{ij}^n=0$
for
$i \neq j$.
To find
$c^n_{ii}$,
let us observe that
$x y^+ \simeq \frac 3 2 T^{01}-m$
and
$x^+ y \simeq \frac 3 2 T^{01}+m$.
Here the sign $\simeq$ is used to denote
the equality up to polynomials of lesser degrees.
Hence, $M^n_{ii} \simeq (3/2)^n
\left( (T^{01} - \frac 2 3 m)^i (T^{01} + \frac 2 3 m)^{n-i} \right)_0$.

It will be proved below that if
$f(m,\, T^{01})$ is an arbitrary polynomial,
then
\be\label{int}
\left( f(m,\, T^{01}) \right)_0 \simeq \frac 1 2 \int_0^1
\left( f(m,\, \frac 2 3 m \tau) + f(m,\, -\frac 2 3 m \tau) \right) d\tau.
\ee
It follows immediately from (\ref{int}) that
\be
M^n_{ii} \simeq \frac 1 2 (\frac 2 3 m)^n \int_0^1
\left( (\tau - 1)^i (\tau + 1)^{n-i} + (-\tau - 1)^i (- \tau + 1)^{n-i}
\right) d\tau
= \nn
\frac {(-1)^{i}} 2 (\frac 2 3 m)^n \int_0^1
\left( (1 - \tau)^i (1 + \tau )^{n-i} + (1 + \tau )^i (1 - \tau)^{n-i}
\right)d\tau \nonumber
\ee
and, as a consequence,
$c^n_{ii}\neq 0 \ (i=0,...,n)$, which, together with
$c_{ij}^n=0$
for
$i \neq j$
yields the non-degeneracy of
$\tilde{M}$  i.e.,  the absence of nontrivial solutions of
(\ref{ur}).

To prove (\ref{int}), it is sufficient to consider the case
$f(m,\, T^{01})=
t^k$, where  $t$ denotes the expression
$T^{01}$.

It follows from the sequence of the obvious equalities
\be
0 &=& \left( \left[ T^{11},\, \left[ T^{00},\,t^k \right]\right]\right)_0
\simeq \left( \left[ T^{11},\, 2k T^{00} t^{k-1} \right]\right)_0  \nn
&\simeq& \left( -4k(k-1)t^{k-2}T^{00}T^{11} -8 k t^k \right)_0
\simeq \left( -4k(k-1)t^{k-2}(9t^2 - 4 m^2) - 8 k t^k \right)_0      \nonumber
\ee
that
\be
\left( t^k \right)_0 \simeq
\frac 4 9 m^2 \frac {k-1}{k+1} \left( t^{k-2}\right)_0
\simeq \cases{
          \displaystyle{\frac 1 {k+1} \left(\frac 2 3 m\right)^k = \int_0^1
                 \left(\frac 2 3 m \tau \right)^k d\tau}
                    & if $k$ is even, \cr
                {}&{}\cr
               0 & if $k$ is odd,  \cr } \nonumber
\ee
which finishes the proof.

\vskip 5 mm
\vskip 5 mm
\noindent {\bf Acknowledgments} \vskip 3 mm
\noindent
The author is
very grateful to M.~Vasiliev for useful discussions.

\end{document}